\documentclass[twocolumn]{aastex631}

\begin{document}


\title{Application of Physics-Informed Neural Networks in Removing Telescope Beam Effects}

\author{Shulei Ni}
\affiliation{Research Center for Astronomical Computing, 
 Zhejiang Laboratory, Hangzhou 311121, China}
 
\author{Yisheng Qiu}
\affiliation{Research Center for Astronomical Computing, 
 Zhejiang Laboratory, Hangzhou 311121, China}

\author{Yunchuan Chen}
\affiliation{Research Center for Astronomical Computing, 
 Zhejiang Laboratory, Hangzhou 311121, China}

\author{Zihao Song}
\affiliation{Research Center for Astronomical Computing, 
 Zhejiang Laboratory, Hangzhou 311121, China}

\author{Hao Chen}
\affiliation{Research Center for Astronomical Computing, 
 Zhejiang Laboratory, Hangzhou 311121, China}

\author{Xuejian Jiang}
\affiliation{Research Center for Astronomical Computing, 
 Zhejiang Laboratory, Hangzhou 311121, China}

\author{Di Li}
\affiliation{Research Center for Astronomical Computing, Zhejiang Laboratory, Hangzhou 311121, China}
 \affiliation{New Cornerstone Science Laboratory, Department of Astronomy, Tsinghua University, Beijing 100084, China}
 \affiliation{National Astronomical Observatories, CAS, Beijing 100012, People's Republic of China}

\author{Donghui Quan}
\affiliation{Research Center for Astronomical Computing, 
 Zhejiang Laboratory, Hangzhou 311121, China}

\author{Huaxi Chen}
\affiliation{Research Center for Astronomical Computing, 
 Zhejiang Laboratory, Hangzhou 311121, China}

\correspondingauthor{Huaxi Chen}
\email{chenhuaxi@zhejianglab.com}






\begin{abstract}
This study introduces {\tt{PI-AstroDeconv}}, a physics-informed semi-supervised learning method specifically designed for removing beam effects in astronomical telescope observation systems. The method utilizes an encoder-decoder network architecture and combines the telescope's point spread function or beam as prior information, while integrating fast Fourier transform accelerated convolution techniques into the deep learning network. This enables effective removal of beam effects from astronomical observation images. {\tt{PI-AstroDeconv}}  can handle multiple PSFs or beams, tolerate imprecise measurements to some extent, and significantly improve the efficiency and accuracy of image deconvolution. Therefore, this architecture is particularly suitable for astronomical data processing that does not rely on annotated data. To validate the reliability of the architecture, we used the SKA Science Data Challenge 3a datasets and compared it with the $\tt{CLEAN}$ deconvolution method at the 21-cm power spectrum level.  The results demonstrate that our algorithm not only restores details and reduces blurriness in celestial images at the pixel level but also more accurately recovers the true neutral hydrogen power spectrum at the power spectrum level.

\end{abstract}



\section{Introduction} \label{sec:intro}
		In the universe, there are various types of astronomical signals originating from the electromagnetic radiation emitted by celestial bodies, systems, or cosmic phenomena~\citep{rohlfs2013tools,Gudel:2002ke}. These signals reveal the nature and characteristics of the cosmic space and are crucial sources of information for humanity's exploration of the universe. Examples of such signals are numerous, including cosmic microwave background radiation, radiation from galaxies and galaxy clusters, and starlight~\citep{Boomerang:2000efg,MARTIN198361,davis1985evolution,Springel:2006vs}. Observing these different types of astronomical signals often requires specially designed telescopes, such as radio telescopes, infrared telescopes, optical telescopes, and X-ray telescopes. The beam width of a telescope is proportional to the wavelength it observes; that is, the longer the wavelength, the wider the beam. However, a beam that is too wide can affect the image clarity and detail resolution. Therefore, different telescope designs and technologies are needed to optimize observational results. Although radio telescopes have a relatively wide beam, their longer wavelengths provide stronger penetration capabilities, allowing them to observe interstellar dust and gas through the Earth's atmosphere. Similarly, due to their longer wavelengths, radio telescopes can expand the observational area through large antenna arrays and improve spatial resolution. Consequently, radio telescopes play a crucial role in detecting and collecting radio waves from celestial radiation, making them ideal tools for exploring distant and mysterious cosmic phenomena such as black holes and dark matter~\citep{schwab1984relaxing,burns2012probing}.

		In recent years, the field of astronomy has achieved numerous breakthroughs and advancements, largely due to the rapid development of astronomical telescopes and related technologies. Among these, radio astronomy has seen significant success, with major radio telescopes such as FAST~\citep{Nan:2011um,li_nan_pan_2012,8331324}, the Arecibo Observatory~\citep{drake1973interstellar,THESTAFFATTHENATIONALASTRONOMYANDIONOSPHERECENTER1975462,hgss-4-19-2013}, the Very Large Array~\citep{napier1983very,perley2011expanded,kellermann2020open}, MeerKAT~\citep{Camilo:2018hsu,10.1093/mnras/sty2785}, CHIME~\citep{Bandura:2014gwa,CHIMEFRB:2018mlh}, LOFAR~\citep{6051244,ROTTGERING2003405,5109710}, ASKAP~\citep{ASKAP:2007rlq,Johnston:2008hp}, and MWA~\citep{tingay2013murchison,5164979} playing crucial roles in advancing our understanding of the universe. The progress in radio astronomy is closely related to the capabilities of aperture synthesis arrays, which form images by cross-correlating signals from a series of antennas. Traditional imaging assumes a uniform primary beam pattern at all locations. MeerKAT, ASKAP, and MWA are precursor projects to the Square Kilometre Array (SKA)~\citep{Santos:2015gra,Bull:2014rha,SKA:2018ckk}, an international radio telescope construction initiative. Many telescopes are expected to be incorporated into the SKA project. With its high sensitivity and resolution, the SKA is set to provide unparalleled observational data.
		
		In cosmology, the Epoch of Reionization (EoR) is a pivotal phase when the universe transitioned from the dark ages to the modern era. During this period, neutral hydrogen (HI) pervaded the entire universe. Due to its hyperfine structure, HI atoms emit radiation at a wavelength of 21 centimeters, hence this characteristic spectral line is known as the 21-cm line. Meanwhile, radiation from early stars, galaxies, and quasars ionized this HI, forming ionized bubbles. These ionized bubbles are crucial for understanding cosmic structure formation and evolution~\citep{datta2009optimal,valdes2013nature,pritchard201221}. The distribution of HI is shaped by cosmic structure formation, and its observational data can significantly constrain cosmological models~\citep{evoli2014unveiling}. By observing and analyzing HI, scientists can gain insights into the universe's transition from the dark ages to the modern era, and explore key issues such as galaxy formation, cosmic structure evolution, and reionization mechanisms. Additionally, the HI signal may carry signatures of dark matter annihilation, offering new prospects for indirect dark matter detection~\citep{gnedin2022modeling}.
		
		However, as the world's largest future synthetic aperture radio telescope system, the SKA faces significant challenges in imaging, including vast amounts of data, complex data management requirements, and the diversity of primary beam patterns~\citep{Wu:2015mq,Wang:2019xwz,Spinelli:2021emp}. The size of the antenna beam, which is the sensitive area for receiving radio waves, affects the imaging quality between antennas, especially in continuous wave interferometry for celestial imaging. The resolution of celestial imaging is determined jointly by the distance between antennas and the size of the beam. Given the large number of antennas in the SKA, such as ASKAP in the central-western region of Australia, which consists of 36 dish antennas with baselines up to 6 kilometers, and MeerKAT in the Northern Cape Province of South Africa, which consists of 64 antennas with baselines up to 8 kilometers, the beam size of these telescopes significantly impacts imaging quality~\citep{Alexander:2011+0,Williamson:2021jrq,Mort:2016,Dodson_2022}. Interferometer baselines can be significantly larger compared to single-dish telescopes, and smaller beams can provide higher spatial and observational resolution, allowing them to detect weaker astronomical signals. However, smaller beams, due to their limited radio wave coverage, lead to a decrease in signal strength, which affects data quality. To achieve sufficient signal-to-noise ratio, data collected through the SKA will require longer integration times or more advanced data processing techniques, ultimately affecting imaging quality. Therefore, researchers must carefully consider the impact of antenna beam patterns in the SKA when conducting research and developing observations.
		
		In the field of data processing for radio telescopes, the {\tt{CLEAN}} deconvolution algorithm plays a crucial role. Since its introduction by Högbom in 1974~\citep{Hogbom:1974jql}, it has not only revolutionized the technology of radio synthesis imaging but also significantly accelerated the development of the entire fielde~\citep{Cornwell:2008zn,Rich:2008sf}. The efficiency of the {\tt{CLEAN}} algorithm stems from its ability to identify and mitigate the effects of point source components in the point spread function (PSF), thereby accurately reconstructing the brightness distribution of the sky~\citep{cornwell1983method}. The algorithm is particularly effective for celestial objects that can be approximated with a small number of parameters, which typically appear as a few widely distributed point sources in the imaging field. However, the {\tt{CLEAN}} algorithm also has some notable drawbacks. For example, its deconvolution performance is less effective for extended sources with continuous intensity distributions. Additionally, when processing large amounts of interferometer data, the {\tt{CLEAN}} algorithm requires substantial computational resources and has relatively slow processing speeds. Moreover, the algorithm relies on researchers having a solid background in astronomical data processing to set the parameters appropriately. Improper parameter settings can lead to underfitting or overfitting issues. With the continuous advancement of radio telescope technology, there is an increasing demand for more complex deconvolution algorithms. To meet this demand, various derivative versions of the {\tt{CLEAN}} algorithm have emerged, such as $\tt{Clark}$~\citep{clark1980efficient}, $\tt{Cotton-Schwab}$~\citep{schwab1983global, schwab1984relaxing}, $\tt{MS-CLEAN}$~\citep{wakker1988multi, cornwell2008multiscale}, and  {\tt{WSClean}}\footnote{\url{https://gitlab.com/aroffringa/wsclean}}~\citep{offringa-wsclean-2014, offringa-wsclean-2017, vandertol-2018}. These derivative algorithms retain the advantages of the {\tt{CLEAN}} algorithm while further enhancing the capability to process complex astronomical data.

		In recent years, with the gradual advancement of artificial intelligence technology, deep learning algorithms have been applied to address issues such as beam effects in radio telescopes~\citep{an2019deep,Ni:2022kxn,schmidt2022deep,chiche2023deep}. Generally, deep learning algorithms have a strong dependence on the quantity and quality of training data and labeled data, which is particularly crucial in the field of astronomy where data labeling is essential. Historically, the application of deep learning algorithms in astronomy has been limited due to the difficulty in obtaining high-quality data. To address this issue and facilitate image deconvolution and reconstruction, we have introduced an unsupervised network architecture that integrates physical prior information, named $\tt{PI-DeconvAstro}$~\citep{ni2024pi}. This network employs an encoder-decoder structure and utilizes the telescope’s point spread function (PSF) or beam as prior knowledge. We tested the network using the SKA Science Data Challenge 3a~(SDC3a)~\footnote{\url{https://sdc3.skao.int/challenges/foregrounds}} datasets and compared the algorithm with the {\tt{CLEAN}} deconvolution algorithm at the level of the power spectrum.

		In studies aiming to reconstruct cosmic evolution history and measure the 21-cm power spectrum using the HI signal, the primary technical challenge lies in effectively mitigating foreground contamination originating from both the Milky Way and extragalactic sources. This foreground contamination, predominantly composed of Galactic synchrotron emission, free-free radiation from interstellar media, and continuous radiation from extragalactic point sources, exhibits an intensity exceeding that of the HI signal by $4\sim5$ orders of magnitude~\citep{Chapman:2014sfa}. Under these circumstances, precise removal of beam effects in observational data becomes critical for data processing. Particularly for the PSF effects of extragalactic point sources, any incomplete or erroneous correction procedures may introduce systematic biases during subsequent signal extraction, ultimately leading to significant distortions in the reconstructed HI power spectrum~\citep{ni2024pi}.

		The structure of the remainder of this paper is organized as follows: Section~\ref{sec:simulation} introduces the data simulation and pre-processing procedures. Section~\ref{sec:deconv} provides a detailed explanation of the deconvolution processes of the {\tt{CLEAN}} algorithm and the {\tt{PI-AstroDeconv}} algorithm. Section~\ref{sec:results} first employs {\tt{PCA}} to remove foreground signals, followed by a comparative analysis of the power spectra obtained using different methods. Finally, Section~\ref{sec:conclusion} presents a comprehensive summary of the entire study.

		\section{Data Simulation and Pre-Processing}\label{sec:simulation}
		In radio astronomy, interferometric array technology is a crucial observational method that significantly enhances the sensitivity and resolution for detecting radiation signals from celestial bodies through the coordinated observations of multiple antenna elements. However, in actual observations, data are often affected by various factors, such as the telescope's beam effects and interference from the Milky Way and extragalactic sources, which can introduce errors and noise into the observational results~\citep{Makinen:2020gvh, Ni:2022kxn, Gao:2022xdb}. To extract valuable astronomical information from raw observational data, meticulous data simulation and pre-processing are required. This chapter delves into this process, providing a detailed description of the data simulation specifics for SDC3a and conducting simulations based on the relevant parameters of SDC3a. 
		
		\subsection{Telescope Model and Foreground Model}
		The principle of aperture synthesis technology involves combining signals from multiple antennas using interferometric techniques to achieve high-resolution image reconstruction~\citep{hogbom1974aperture}. The key is to arrange the telescopes in an array and observe the same light source simultaneously~\citep{Wilson2009,thompson2017interferometry}. Each telescope receives signals that carry phase information, reflecting the spatial structure of the source. In aperture synthesis, the first step is to process these signals for correlation, which is typically achieved through Fourier transformation. Fourier transformation converts signals from the time domain to the frequency domain, allowing for the measurement and recording of the phase difference between antennas. By measuring the amplitude and phase of the correlated signals and performing appropriate data processing and analysis, high-resolution images can be reconstructed\citep{Thompson:2001ms}.
		
		The SDC3a simulation project represents the third Science Data Challenge of the SKA radio telescope, with its core objective being the effective separation of foreground contamination signals from the Milky Way and extra-galactic sources to reveal the pristine HI signal. In the SDC3a simulation, the $\tt{OSKAR}$~\footnote{\url{https://github.com/OxfordSKA/OSKAR}} software package is employed to generate visibilities, which are produced based on specific time and frequency sampling~\citep{hamaker2006understanding,mort2010oskar,Dewdney:2017}. The telescope model is based on the SKA-Low configuration, simulating the ``Vogel'' layout that includes 512 stations, a single-arm spiral configuration designed to achieve uniform antenna area density and maximized azimuthal angle sampling. To reduce computational costs, a single station layout, comprised of 128 antennas, was adopted in the simulation rather than $512$ distinct station layouts. This simplification resulted in an increased response in the far side-lobes, for which compensation was made in the definition of the foreground model.
		
		The foreground model consists of two parts: the ``outer'' component covering a $2\pi$ steradian volume above the horizon and the ``inner'' component defining a more limited field of view near the pointing direction. The external component includes the A-Team sources with brightness exceeding $1.48\sim2.83\times10^5\mathrm{K}$ and sources from the $\tt{GLEAM}$ catalogue for which a composite $\tt{GLEAM}$ and $\tt{LoBES}$ $\tt{OSKAR}$ model file was kindly made available for use in these simulations~\citep{Wayth:2015nla,Lynch:2021xzb}. All sources with a brightness exceeding $9.32\times10^3~\mathrm{K}$ at $150~\mathrm{MHz}$ (approximately $1200$ sources) are included in the external sky model. These sources, located in the far side-lobe regions of the observation, are typically modeled and removed through a ``deconvolution'' process during the observation and imaging procedures. To simulate the attenuation of these sources, a net attenuation factor of $10^{-3}$ is assumed. 
		
		The inner foregound model is an integral component of radio astronomical simulations, accurately depicting the distribution of sky sources at the observational frequency. This model is anchored on the first null of the station beam pattern at the minimum observational frequency, ensuring high-fidelity in the simulation. The model encompasses all sources from the $\tt{GLEAM}$ and $\tt{LoBES}$ catalogues with a $150~\mathrm{MHz}$ brightness exceeding $185.36~\mathrm{K}$, spatially represented through Gaussian models. For sources with a brightness below $1.8\times10^5~\mathrm{K}$ , the model employs the $\tt{T-RECS}$~\footnote{\url{https://github.com/elucherini/t-recs}} code for simulation, extending down to a brightness of $1.864 \mathrm{K}$, over an $8\times8~\mathrm{degree}$ simulation area~\citep{Lucherini2021TRECSAS}. The model utilizes a $5\times5~\mathrm{arcsec}$ pixel sampling with a $10\times10$ $\mathrm{arcsec}$ FWHM Gaussian gridding kernel to normalize the data to units of $\mathrm{K/pixel}$, facilitating analysis and comparison. Within the model, pixels brighter than $14.8~\mathrm{mK/pixel}$  are considered as Gaussian source model components with a $10\times10~\mathrm{arcsec}$  extent, having a flux density equivalent to the pixel's brightness, and a spectral index set to zero. The model is also updated for each observational frequency to track the spectral energy density variations of $\tt{T-RECS}$ sources, adapting to the characteristics of sources under different observational conditions. This detailed inner sky simulation provides a robust foundation for radio astronomical data processing, especially in dealing with beam effects and foreground interference.
		
		The inner foreground model's additional components were created using a coarser $512\times512$ pixel grid with $56.25\times56.25$ arcsec sampling, balancing accuracy with computational demands. This resolution matches the $\tt{GSM2016}$~\citep{Zheng:2016lul} sky model's evaluation at various observing frequencies. To address $\tt{GSM2016}$'s limited angular resolution at low radio frequencies, the model was enhanced with $\tt{MHD}$ simulation data~\citep{Sault:1995vp,Jelic:2008jg}, capturing filamentary structures down to the $\mathrm{arcmin}$ scale and scaled to $\mathrm{K/pixel}$. Furthermore, faint extra-galactic sources from the residual $\tt{T-RECS}$ image were smoothed and resampled to fit the ``inner'' foreground model grid, using appropriate normalization for the coarser sampling. This approach ensures a comprehensive representation of the sky for radio astronomical simulations.
		
		\subsection{EoR Signal and Error Model}
		The signal data from the EoR is derived from $\tt{21cmFAST}$~\footnote{\url{https://github.com/21cmfast/21cmFAST}} simulations. This simulation technique can generate three-dimensional cosmic evolution scenarios, including detailed representations of density, ionization fraction, peculiar velocity, and spin temperature fields. These integrated fields are further used to derive the brightness temperature distribution at a $21~\mathrm{cm}$ wavelength~\citep{Mesinger:2010ne,Park:2018ljd,Murray:2020trn,Munoz:2021psm}. The brightness temperature of HI is a key physical quantity that describes the intensity of its 21-cm hyperfine transition radiation. It is typically expressed as:
		\begin{equation}\label{temp}
			T_b = (T_s - T_{cmb})(1-e^{-\tau}) + T_{cmb},
		\end{equation}
		where  $T_s$ denotes the spin temperature characterizing the distribution of HI atoms in the ground state hyperfine levels, $T_{cmb}$ represents the temperature of the CMB, $\tau$ is the optical depth for absorption through the cloud at a given frequency. The $\tt{21cmFAST}$ ensures swift prediction of the 21-cm signal while maintaining simulation speed and efficiency, enabling efficient computation on a single processor~\citep{Mesinger:2010ne}.
		
		In the process of the HI simulation, a set of fixed cosmological parameters were selected, which are based on the best-fit values obtained from the $\it{Planck}$ 2018 data~\citep{Planck:2018vyg}. The CMB temperature is $T_{\text{CMB}} = 2.7255 \, \text{K}$. The matter density is $\Omega_{\text{m}} = 0.30964$, while the dark energy density is $\Omega_{\Lambda} = 0.69036$. The Hubble constant is given by $H_0 = 100 \, \text{km}\, \text{s}^{-1}\, \text{Mpc}^{-1}$\footnote{To avoid unit conversion issues in calculating the 21-cm power spectrum, the organizers of SDC3a adopt $H_0 = 100 \, \text{km}\, \text{s}^{-1}\, \text{Mpc}^{-1}$.}. The effective number of neutrino species is $N_{\text{eff}} = 3.046$, with three neutrino masses specified as $m_{\text{nu}} = [0, 0, 0.06] \, \text{eV}$. The baryon density is $\Omega_{\text{b}} = 0.04897$.

		To enhance the realism of the simulation, SDC3a incorporated various instrumental noises into the simulation process. They artificially attenuated the brightness of sources in the ``outer" sky model beyond the central 8x8 degrees with a factor of $10^{-3}$, simulating a partially successful full-sky source population modeling and subtraction. This attenuation reflects the differences in far side-lobe responses due to diverse station layouts and the precision of source modeling and subtraction~\citep{Dewdney:2017}.
		
		\subsection{Simulated Images}
		During the simulation process, the $\tt{OSKAR}$ software utilizes specific telescope, sky, and error models to generate visibility data. Images are then created from this visibility data along with their corresponding synthesized beams. Once the images and the PSF for all frequency channels are defined, they are assembled into a data cube for subsequent analysis.

		During the simulation process, the signal, foreground, and noise components were represented as $N\times512\times512$ pixel cubes, where $N$ denotes the number of frequency channels. However, when extracting the visibility function, we skillfully employed the oversampling parameter to optimize the output to an $N\times2048\times2048$ pixels cube. This strategy significantly improved the precision of data processing, resulting in more accurate images~\citep{hjellming1989design}.
		
		If only a small region of the sky needs to be mapped, the visibility $V(u, v, w)$ can be represented as follows~\citep{Thompson:2001ms}:
		\begin{equation}\label{eq:visi}
			V(u, v, w)=\int_{-\infty}^{+\infty}\int_{-\infty}^{+\infty}A(x, y)I(x, y)e^{i2\pi(ux+vy+w)}dxdy,
		\end{equation}
		where $A(x, y)$ is primary beam, representing the effective collecting area, on a plane perpendicular to the x and y directions, and $I(x, y)$ is intensity distribution.
        
        With the flat sky approximation, one can approximate the observed phase $V(u, v, w)$ to the $uv$-plane, $V(u, v, w)\simeq V(u, v, 0)e^{i2\pi w}$. Then, perform the inverse Fourier transform on Equation~(\ref{eq:visi}), and we can obtain
		\begin{equation}\label{eq:Ip_xy}
			I'(x, y) = A(x, y)I(x, y)=\int_{-\infty}^{+\infty} V(u, v, 0)e^{-i2\pi (ux+vy)}dudv,
		\end{equation}
		where $I'(x, y)$ is the intensity $I(x, y)$ as modified by the primary beam function $A(x, y)$. One can easily correct $I'(x, y)$ by dividing it, pixel by pixel, by $A(x, y)$. By solving Equation~(\ref{eq:Ip_xy}), the aperture synthesis can be derived from the varying $I(x, y)$ values among different sources. However, this necessitates a substantial computational capacity. To simplify the problem, we have to grid the visibility function and weight it by a grading function, $g(u, v)$.  We can obtain an image via a discrete Fourier transform (DFT):
		\begin{equation}\label{eq:ID_xy}
			I_D(x, y)=\sum_k g(u_k, v_k)V(u_k, v_k)e^{-i2\pi(u_kx+v_ky)}.
		\end{equation}
		Images are then created from this visibility data along with their corresponding synthesized beams.
		Comparing Equation~(\ref{eq:Ip_xy}) and Equation~(\ref{eq:ID_xy}), we can derive the dirty image, corresponding to the following
		\begin{equation}\label{eq:ID}
			I_D(x, y) = P_D(x, y)\otimes I'(x, y),
		\end{equation}
		where $\otimes$ means convolution, and 
		\begin{equation}
			P_D=\sum_k g(u_k, v_k)e^{-i2\pi(u_kx+v_ky)}.
		\end{equation}
		The $P_D$ is response to a point source, which is is known as the dirty beam or the PSF. It is worth noting that in the subsequent construction of the deep learning network, we will innovatively introduce $P_D$ as prior physical information and deeply integrate it into the network architecture. This strategy aims to leverage the guiding role of physical knowledge to achieve deconvolution while ensuring the interpretability and accuracy of the results.
		
		Since we are not concerned with the weighting method but rather with examining the effects of deconvolution, we use ``natural'' weighting, $g(u, v)=1$ throughout this work. This approach avoids assigning weights to samples before gridding, ensuring that regions with more baselines do not dominate the final image formation. This method avoided assigning weights to samples prior to gridding, ensuring that regions with more baselines did not dominate the formation of the final image.
		
		The specific characteristics of the data have been meticulously documented in Table~\ref{tab:data}, which includes parameters such as the observation track length ($\text{HA}$), thermal noise equivalent (${\text{N}}_{\text{thermal}}$), field of view ($\text{FoV}$), integration time ($\text{Ti}$), channel width ($\Delta \text{f}$), frequency coverage (${\text{f}}_{\text{cov}}$), Pixel size ($\text{px}$), and Number of pixels (${\text{N}}_{\text{pix}}$). 
  
        Additionally, the SDC3a simulation incorporates various ancillary data, including synthesized beam and time-varying station beams at each frequency, station and antenna layouts, and thermal noise models. These detailed parameter settings facilitate subsequent analysis and discussion. Table~\ref{tab:data} not only encapsulates the key parameters of the dataset but also provides essential insights for further exploration of the data's properties. Through a rigorous examination of these data features, our aim is to attain a more profound comprehension of the simulation outcomes and to offer robust foundational support for future research endeavors.

        \begin{table*}\caption{General parameters, measurement set configurations, and image cube specifications for the \texttt{21cmFAST} simulation.}
            \centering
            \begin{tabular}{|c|c|c|c|c|c|c|c|c|}
                \hline
                \textbf{Parameters} & \textbf{HA} & \boldmath$N_{\textbf{thermal}}$ & \textbf{FoV (RA, Dec)} & \textbf{Ti} & \boldmath$f_{\textbf{conv}}$ & \boldmath$\Delta f$ & \boldmath$N_{\textbf{pix}}$ & \textbf{pix} \\
                \textbf{[Unit]} & [Hours] & [Hours] & [deg] & [sec] & [$\mathrm{MHz}$] & [$\mathrm{KHz}$] & [No.] & [arcsec] \\
                \hline
                \textbf{Values} & $[-2, 2]$ & $1000$ & $[-30, 0]$ & $10$ & $106\text{--}196$ & $100$ & $512{\times}512$ & $16{\times}16$ \\
                \hline
            \end{tabular}
            
            \label{tab:data}
        \end{table*}

\begin{figure}
			\centering
			\includegraphics[width=0.3\textwidth]{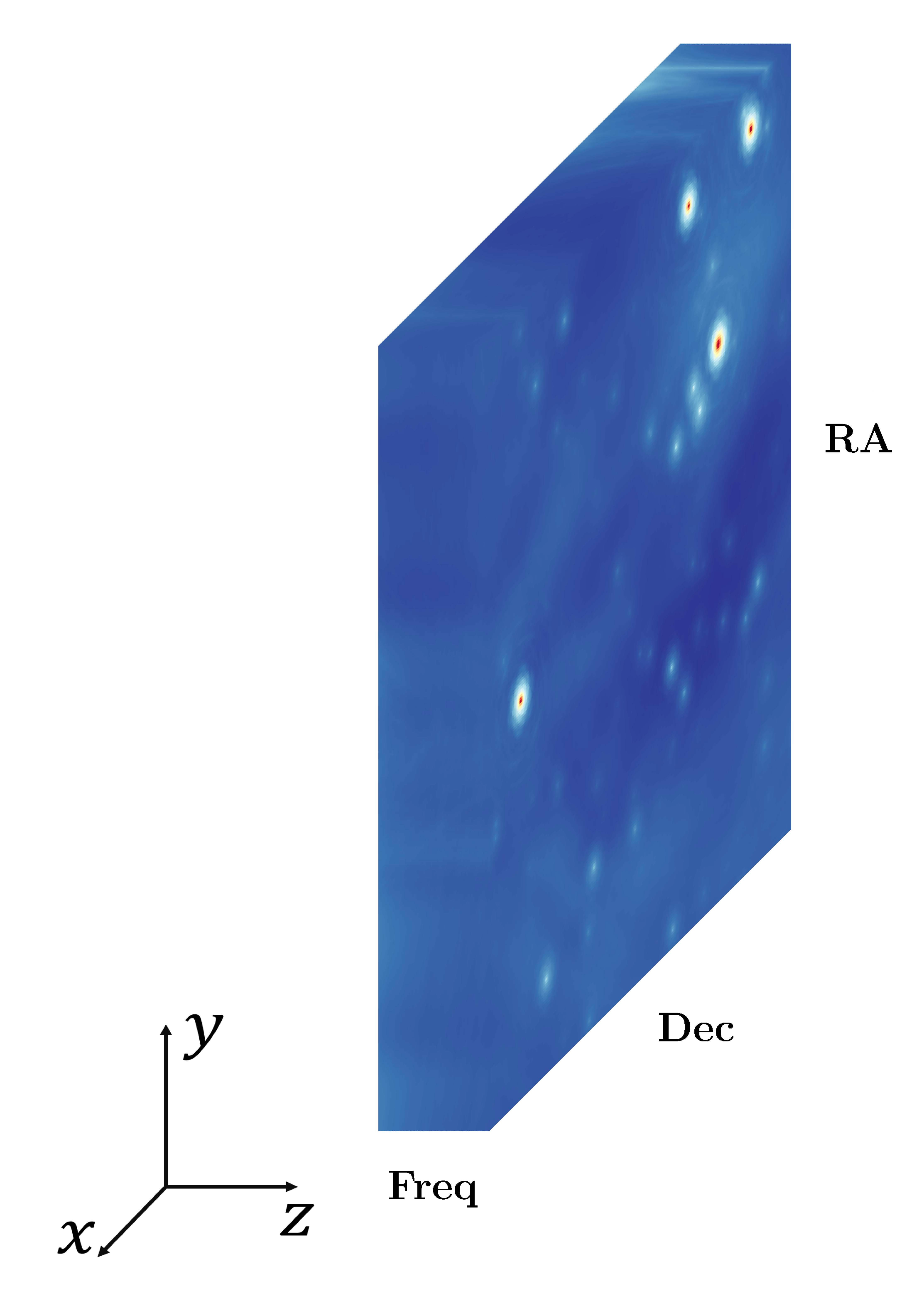}
			\caption{SDC3a simulation image schematic. The $xy$-plane represents scale, with the $x$-axis indicating Declination and the $y$-axis representing Right Ascension. The $z$-axis denotes the line of sight direction, which corresponds to the frequency direction. The bright regions represent the distribution of galaxy clusters.}
			\label{fig:cube_3d}
		\end{figure}

		To thoroughly investigate the spatial characteristics of the SDC3a simulation data, we have constructed a 3D image, as shown in Figure~\ref{fig:cube_3d}. Within this figure, the $xy$-plane corresponds to the scale distribution in the celestial coordinate system. Specifically, the $x$-axis represents the declination coordinate, the $y$-axis denotes the right ascension coordinate, and the $z$-axis symbolizes the line-of-sight direction, or the frequency axis direction. Given that the core region of the SDC3a simulation is concentrated on the central area of the $xy$-plane, namely the central $512\times512$ pixels of that plane, we have deliberately selected and displayed only the details of this central region in the presented image. The bright regions in the figure represent the distribution of galaxy clusters. 
		
		\subsection{Data Pre-processing}
		
		In the previous subsection, we provided a detailed description of the data parameters and the models used in the SDC3a simulation. Next, we will pre-process the data for training. Given that the ultimate goal of processing SDC3a data is to generate foreground subtracted 21-cm power spectra, each pixel value can significantly influence the magnitude and characteristics of the power spectrum. Therefore, we will not normalize the data to avoid altering its key features, even if normalization simplifies training.
		
		During actual observations, visibility data obtained from interferometric arrays are converted into dirty images and dirty beams using formulas Equation~(\ref{eq:visi}-\ref{eq:ID}). Since our study focuses on evaluating the deconvolution capabilities of the {\tt{PI-AstroDeconv}}  algorithm, we use only {\tt{WSClean}} at this stage to convert visibility data into images without performing any deconvolution operations, i.e. niter=0. Subsequently, the generated dirty images and dirty beams are used as training data and physical prior information, respectively, and are input into the {\tt{PI-AstroDeconv}} network.  To balance computational cost, we selected only the frequencies in the $166–181~\mathrm{MHz}$~(150 frequency slices) band from the SDC3a data for analysis in the comparison with the CLEAN algorithm.
        
        Due to the fact that the beam effect only influences its corresponding specific frequency and has no impact on neighboring frequencies, our network is designed for single-frequency processing, yet it enables simultaneous training across multiple frequencies.
		
		This approach not only preserves the true characteristics of the data but also allows us to more accurately assess the effectiveness of incorporating dirty beams as prior physical information within the network architecture, thereby enhancing the interpretability and accuracy of the results. Our method leverages physical knowledge to inform deconvolution; preventing unnecessary complexity and artifacts. Furthermore, our model ensures physical consistency while significantly improving computational efficiency.

		\section{De-convolution}\label{sec:deconv}
		In the data processing of radio astronomy, deconvolution techniques play a critical role by minimizing the limitations of visibility measurements. The visibility function has two main shortcomings that constrain the accuracy of aperture synthesis imaging: limited spatial frequency coverage and inherent errors in the visibility function. Deconvolution techniques can effectively address these deficiencies.
		
		Based on the {\tt{CLEAN}} algorithm, various deconvolution algorithms have been developed and they perform excellently in astronomical image deconvolution. Deep learning algorithms like {\tt{PI-AstroDeconv}}  offer new approaches to deconvolution. This chapter will primarily focus on the performance and application of {\tt{WSClean}}, as well as the {\tt{PI-AstroDeconv}}  algorithm.
		
		\subsection{{\tt{WSClean}}}
		In the data processing of radio astronomy, due to the incomplete coverage of the $uv$ plane, the resulting dirty images exhibit significant sidelobes and other artifacts, which differ markedly from the true sky map and are unsuitable for direct scientific analysis. To obtain higher-quality images, additional observations can be conducted to improve the coverage of the $uv$ plane, or prior knowledge can be used to interpolate the uncovered regions of the $uv$ plane, which involves deconvolution processing.
		
		{\tt{WSClean}} ({\tt{W-Stack~CLEAN}}) is an advanced imaging tool widely used in radio astronomy for generating high-quality images from interferometric data. Designed to handle the extensive datasets produced by modern radio telescopes, {\tt{WSClean}} employs the w-stacking algorithm to efficiently correct for wide-field effects, such as sky curvature, enabling accurate imaging over large fields of view. It supports multi-scale {\tt{CLEAN}}  and wideband deconvolution, allowing for detailed reconstruction of astronomical sources across multiple frequencies. With its optimized parallel processing capabilities and extensive customization options, {\tt{WSClean}} is ideal for producing high-fidelity images in a variety of observational contexts, from surveys to targeted studies.
		
		The {\tt{CLEAN}}  algorithm assumes that the target source can be represented as a series of point sources. It then performs a Fourier transform of the visibility function to compute the image and a Fourier transform of the weighted spatial transfer function to calculate the point source response. The dirty image and dirty beam are synthesized separately. The algorithm identifies the brightest point source in the image and subtracts the point source response for that location, which is the dirty beam centered at that point and containing all sidelobes. The peak amplitude of the subtracted point source response is set to a fraction (typically one-tenth) of the strength of the point in the image, known as the loop gain. Subsequently, a Dirac function component is inserted into the model to mark the location and amplitude of the detected component. By iteratively repeating this process, a cleaned image is ultimately obtained.

\begin{figure*}
			\centering
			\includegraphics[scale=0.5]{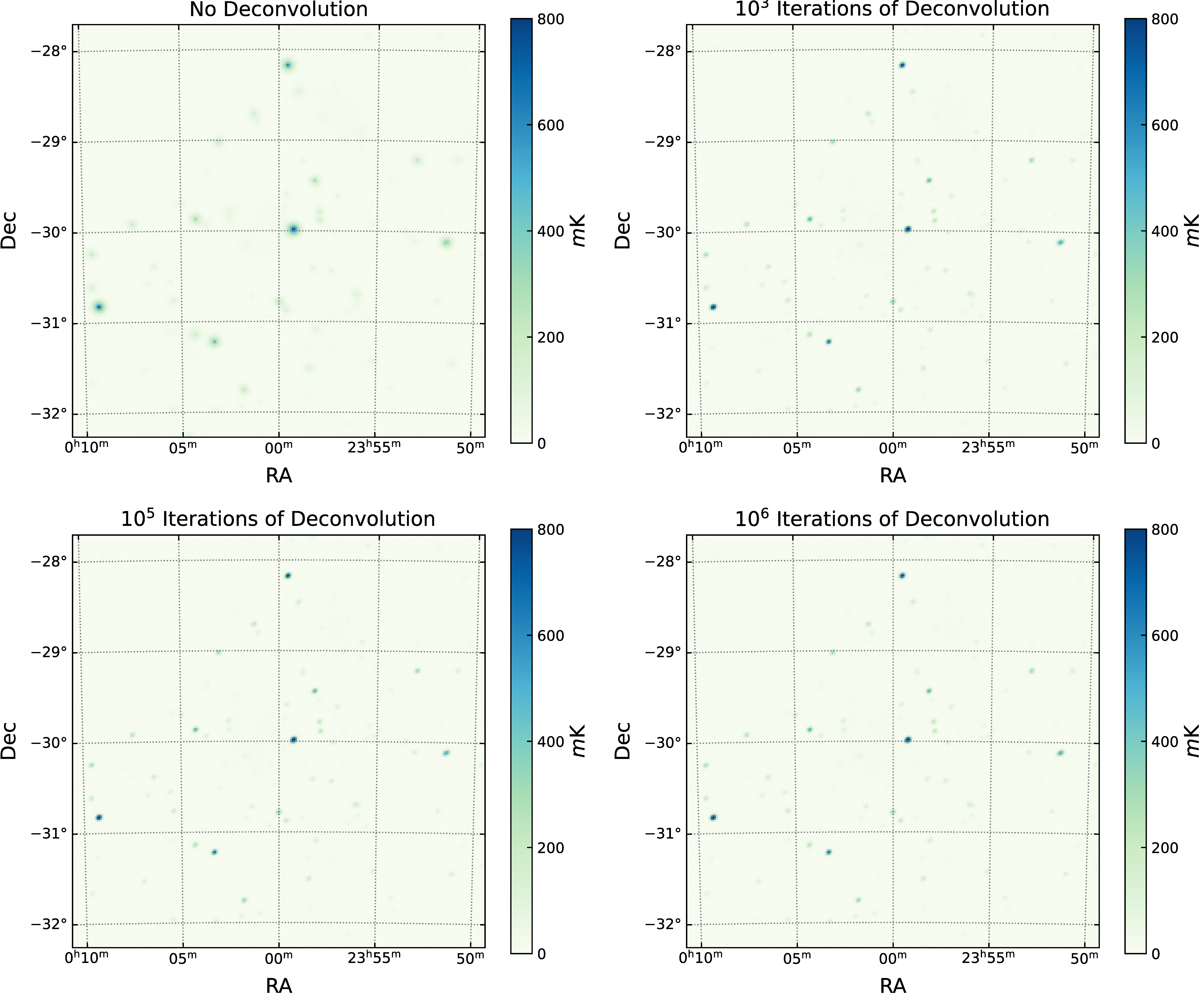}
			\caption{Deconvolved $166~\mathrm{MHz}$ brightness temperature images for iterations. Top-left panel: no deconvolution; Top-right panel: $1~\mathrm{K}$ iterations; Bottom-left panel: $100~\mathrm{K}$ iterations; Bottom-right panel: $1~M$ iterations. The unit in the images is $mK$.}\label{fig:wsclean_result}
		\end{figure*}

		In Figure~\ref{fig:wsclean_result}, we present images generated from 166 $\mathrm{MHz}$ simulated data after varying numbers of deconvolution iterations. The upper left image shows the data without any deconvolution processing; the upper right image corresponds to the result after $1000$ iterations; the lower left image demonstrates the effect of $100,000$ iterations; and the lower right image represents the image after $1,000,000$ iterations. During the processing, we employed natural weighting to ensure a resolution of $16$ arcseconds per pixel and uniformly output all images at a size of $2048\times2048$.
		To enable  {\tt{PI-AstroDeconv}}  fully utilizing the detailed information provided by the telescope, we deliberately retained the PSF corresponding to each image. Naturally, the results exhibit significant variations due to the different numbers of deconvolution iterations. To ensure the comprehensiveness of the network input and maintain the integrity of the physical information contained in the PSF, we performed direct imaging on the visibility function without any convolution operations. This result is presented in the upper left image of Figure~\ref{fig:wsclean_result}.

		In data processing, the {\tt{CLEAN}}  method is a widely used technique for improving the quality of individual radio interferometric images. However, it does not produce a unique and stable solution and requires considerable computational resources. While the {\tt{CLEAN}}  algorithm is relatively easy to understand as a process for removing beam effects, it is a highly nonlinear method and does not provide a complete mathematical analysis.	
		
		When using {\tt{WSClean}}, several parameters must be manually chosen, such as selecting between natural weighting, uniform weighting, or Briggs weighting, which can affect the spatial resolution and noise level of the resulting image. The number of deconvolution iterations impacts the required computation time and result quality, with the number of iterations heavily relying on the user’s data processing experience. Other parameters include loop gain and window size. Additionally, {\tt{WSClean}} is known to introduce spurious structures such as scattered spots or wave-like features on extended source characteristics. To address these issues, we are exploring the use of deep learning algorithms.
		
		\subsection{{\tt{PI-AstroDeconv}}}\label{sec:deconv-unet}
		
		This section provides a comprehensive description of the {\tt{PI-AstroDeconv}}  architecture, including a detailed discussion on the loss functions and training strategies.

		\subsubsection{Architecture}
		This subsection focuses on the network architecture. Our network is constructed based on {\tt{U-Net}}, which was proposed by Ronneberger et al. to address the medical image segmentation problem~\citep{Ronneberger:2015UNetCN,Cicek:2016abs,Isensee:2019AnAA,Makinen:2020gvh,Ni:2022kxn}. {\tt{U-Net}} consists of symmetric encoder and decoder components, where the encoder is used to extract high-level features of the image and the decoder is used to restore the feature maps to the original image size and perform segmentation. The core of the network is the U-shaped structure, which enables the network to capture both local and global information, resulting in more accurate segmentation. Since its inception, the {\tt{U-Net}} network has been widely applied in various fields beyond segmentation problems, including regression problems. Specifically, the encoder is made up of multiple convolutional layers and pooling layers that are used to extract high-level semantic features from the image. The decoder is made up of upsampling layers and convolutional layers, which are used to restore the feature maps to the original image size and perform segmentation. In the decoder, each up-sampling layer is concatenated with the corresponding feature map in the encoder, via skip connections, to exploit more abundant information. The skip connections enable the decoder to utilize the high-level semantic information from the encoder without losing detail information, thereby enhancing the network's expression ability and segmentation performance~\citep{Ronneberger:2015UNetCN}.

\begin{figure*}
			\centering
			\includegraphics[width=0.95\textwidth]{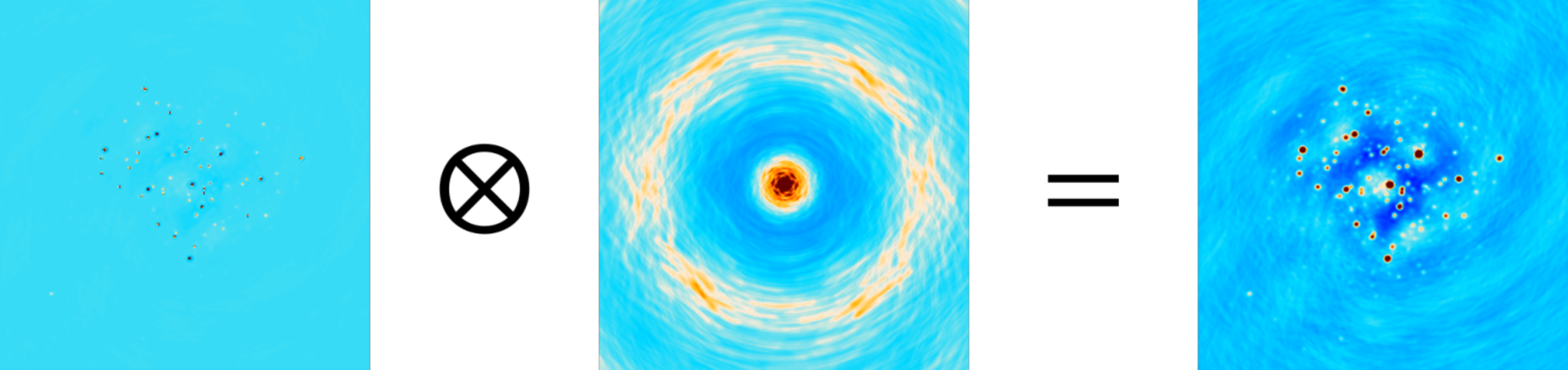}
			\caption{Schematic diagram of telescope observation effects. The image on the left displays the true, clean image of the observed target, while the image in the center reveals the shape of the telescope's PSF or beam. The image on the right illustrates the blurring effect caused by the PSF or beam. The symbol ``$\otimes$'' denotes the convolution operation.}\label{fig:conv_img}
		\end{figure*}

        In Equation~(\ref{eq:ID}), the dirty beam is considered as a filter, playing the role of a convolutional kernel in deep learning~\footnote{Note: The convolution operation in deep learning differs from the conventional concept of convolution.}. To illustrate this process more intuitively, we present a schematic diagram of the sky map convolution in Figure~\ref{fig:conv_img}. The left panel depicts the target clean sky brightness distribution (without beam effects). The middle panel shows the synthesized beam (PSF). The right panel represents the predicted convolved image after applying the beam, which corresponds to the network output used for subsequent loss calculation against observational data. The symbol ``$\otimes$'' represents the convolution operation, and ``$=$'' indicates the equivalence relationship.
		
		Based on the aforementioned theory, we propose a physics-informed unsupervised learning method for astronomical image deconvolution, namely {\tt{PI-AstroDeconv}}, to effectively mitigate the beam effect. Furthermore, considering the substantial size of the dirty beam in aperture synthesis observations, we propose to use the Fast Fourier Transform (FFT) technique, referred to as FFT convolution, in the last layer of {\tt{U-Net}} instead of traditional convolution methods, aiming to improve the computing efficiency. All other layers of the network retain the standard convolutional operations commonly used in deep learning models.

		Our network has undergone some minor modifications compared to standard {\tt{U-Net}} network. A depthwise convolutional layer is added between the {\tt{U-Net}} and the label, and the convolution kernel of this layer is like the beam of a telescope. The input and the label (Ground Truth) in the network are the same, both being observational data from a single telescope. The objective of the network is to produce the same image as the input. The final prediction of the network is the output of the {\tt{U-Net}} before depthwise convolution.
		
\begin{figure*}
			\centering
			\includegraphics[width=0.8\textwidth]{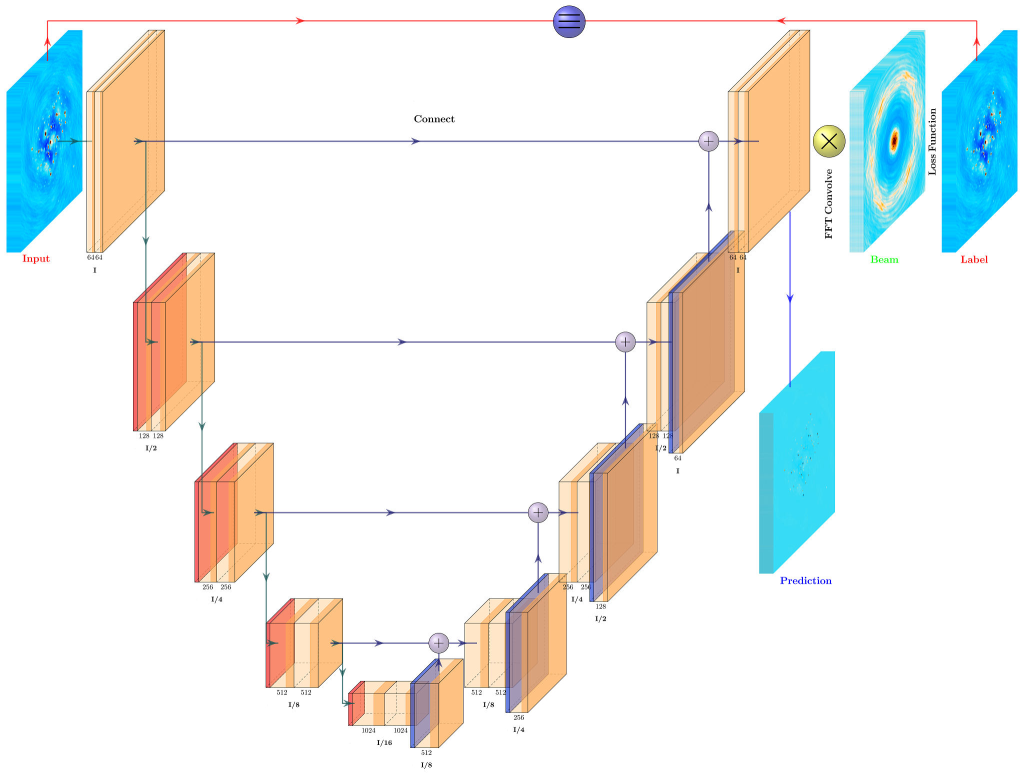}
			\caption{The {\tt{PI-AstroDeconv}}  architecture. The yellow blocks represent convolutional layers, red blocks represent pooling layers, and gray-green blocks represent upsampling operations. The left half of the network depicts the downsampling path, while the right half represents the upsampling path. The number of channels is indicated at the bottom of each block. The arrows above the image symbolize the connections between network layers. The numbers below each layer in the network indicate the quantity of convolutional kernels present in that specific layer. Additionally, the letters denote the variations in the image dimensions within the layer. The letter "I" specifically represents the size of the input image. The blocks located at the far left and far right of the network indicate the input and the label, respectively (represented as deep blue circles). A Depthwise convolutional layer, represented as a yellow circle, is inserted between the {\tt{U-Net}} network and the label. This layer utilizes convolutional kernels resembling telescope beams and is accelerated using FFT. The {\tt{U-Net}}'s output is convolved with the beam, and the result is compared with the labels (Ground Truth) to compute the loss function. Network inputs and ground truth labels are sourced from the telescope. The model is trained to embed physical priors through convolutional layers, ensuring outputs align with the original observations. The final prediction of the network is derived from the last layer of the {\tt{U-Net}}, represented as the last yellow layer. The symbols $+$, $\times$, and $\equiv$ represent concatenation, convolution, and exact equality, respectively. This visualization was created by modifying the {\tt{PlotNeuralNet}} library~(\url{https://github.com/HarisIqbal88/PlotNeuralNet}).}
			\label{fig:deconv_unet}
		\end{figure*}

		The architecture of {\tt{PI-AstroDeconv}}  is illustrated in Fig.~\ref{fig:deconv_unet}. It uses color-coded blocks: yellow for convolutional layers, red for pooling layers, and gray-green for upsampling operations. The left half of the network illustrates the downsampling path, while the right half represents the upsampling path. The number of channels is indicated at the bottom of each block, and the arrows above the image represent the connections between the network layers. The numbers below each layer in the network denote the number of convolutional kernels present in that specific layer. Moreover, the letters indicate variations in image dimensions within the layer, with ``I" specifically denoting the size of the input image. The blocks on the far left and far right of the network represent the input and label, respectively, illustrated as deep blue circles. Between the {\tt{U-Net}} network and the label, a depthwise convolutional layer appears as a yellow circle. This layer utilizes convolutional kernels resembling telescope beams and is accelerated using FFT. Network inputs and ground truth labels are sourced from the telescope. The model is trained to embed physical priors through convolutional layers, ensuring outputs align with the original observations. The network's final prediction emerges from the last layer~(the yellow one) of the {\tt{U-Net}}. The symbols $+$, $\times$, and $\equiv$ represent concatenation, convolution, and exact equality, respectively.
		
		In actual astronomical observations, the dimensions of both the imaging and the telescope‘s PSF are typically large. In order to preserve the integrity of the convolved images and the PSF, we abstained from employing any segmentation and instead directly performed convolution calculations. However, the use of the large convolution kernel ($2048\times2048$) in the final layer of the network hindered the learning process. To address this issue, we implemented a transformation technique that combines Fourier transform and the convolution. More specifically, we replaced the convolution operation in the last layer with Fourier transform, resulting in a reduction of the time complexity from $\mathcal{O}(n^4)$ to $\mathcal{O}(n^2\times log(n))$~\citep{ni2024pi}. Our comparative tests demonstrated that enabling FFT acceleration significantly reduced processing time during training. This improvement exceeded a factor of $10^5$ for large-kernel convolutions in the network's final layer.

		Our objective is to eliminate the beam effects of telescopes through a network design that facilitates the use of data under known beam conditions, without the need for uncontaminated data as ground-truth labels. Additionally, we can replace  {\tt{U-Net}} (shown as the pink block in Figure~\ref{fig:deconv_unet}) with any other network architecture and use the beam of any telescope (shown as the Beam block in Figure~\ref{fig:deconv_unet}). Thus, broadly speaking, our architecture can be regarded as a universal unsupervised framework for eliminating telescope beam effects.".
		
		\subsubsection{Loss Function}
		Our objective is to utilize supervised regression algorithms for the purpose of mitigating the adverse beam effects present in the data obtained from telescope observations. The {\tt{PI-AstroDeconv}} architecture employs input values to predict continuous outputs. When selecting a loss function, it is crucial to ensure the continuity and differentiability of the information. We have experimented with several regression loss functions, such as the mean absolute error (MAE) (L1 norm), mean squared error (MSE) (L2 norm), Huber, and Log-Cosh~\citep{wang2020comprehensive}.
		
		Loss functions vary in continuity and differentiability. MAE is not differentiable at zero, leading to potential discontinuities in optimization. MSE is continuous and differentiable everywhere, with its derivative being twice the error, ensuring a smooth optimization process but is sensitive to outliers, risking overfitting. The Huber loss acts like MSE for small errors ($< \delta$) and MAE for larger errors, differentiable for small errors but not at $\delta$. Log-Cosh loss is continuous and twice-differentiable everywhere, with a derivative expressed as a function of error, providing a very smooth optimization process~\citep{wang2020comprehensive}.

		Considering the robustness to outliers and the second-order differentiability offered by the Log-Cosh loss function, we lean towards regarding it as the optimal choice. 
		\begin{equation}\label{equ:loss_fun}
			L(p, t)=\sum_i\log\cosh(p_i-t_i),
		\end{equation}
		In the case of minor losses, the Log-Cosh function exhibits similarities to the MAE function, whereas for substantial losses, it mirrors the MSE function while retaining its second-order differentiability. Conversely, the Huber loss function lacks differentiability in all scenarios. The MAE loss represents the average absolute error and solely considers the mean absolute distance between the predicted and expected data, rendering it incapable of addressing significant prediction errors. On the other hand, the MSE loss emphasizes crucial errors through squaring, thereby significantly influencing performance metrics. Thus, owing to the Log-Cosh function's commendable robustness to outliers, we designate it as the most fitting approach.
		
		\begin{table}
			\centering
			\caption{Description of the hyperparameters in the {\tt{U-Net}} architecture design.}
			\begin{tabular}{llcc}
				\hline\hline
				Hyperpara    & Description                            & Prior                    & Optimum\\
				\hline
				$\omega$          & weight decay             & [$10^{-4}, 10^{-5}, 10^{-6}$] &$10^{-5}$   \\
				$n_{\rm filters}$ & filters  & [16, 32]                            & 32   \\
				$b$               & batch size  & [1, 2, 4]  & 4   \\
				$\Omega$          & optimizer & [Adam, NAdam]     & NAdam   \\
				\hline
			\end{tabular}\label{tab:unet}
		\end{table}

		\subsubsection{Training Strategy}
		According to the reaches of \cite{Makinen:2020gvh} and \cite{Ni:2022kxn}, and considering the memory limitations of the GPU used, we primarily focus on two sizes of convolutional kernels, $32$ and $16$, at the beginning of the input. The size of the convolutional kernel determines the field of view for convolution, and the sizes under consideration are $5\times5$ and $3\times3$. To achieve the desired output dimension, we employ ``same'' padding for both convolution and transpose convolution to manage the boundaries of the samples. The stride parameter determines the step size at which the convolutional kernel traverses the image. In our model, we preserve the default settings, where the convolutional stride is $1$ and the transpose convolution stride is $2$.
		
\begin{figure}
			\centering
			\includegraphics[width=0.4\textwidth]{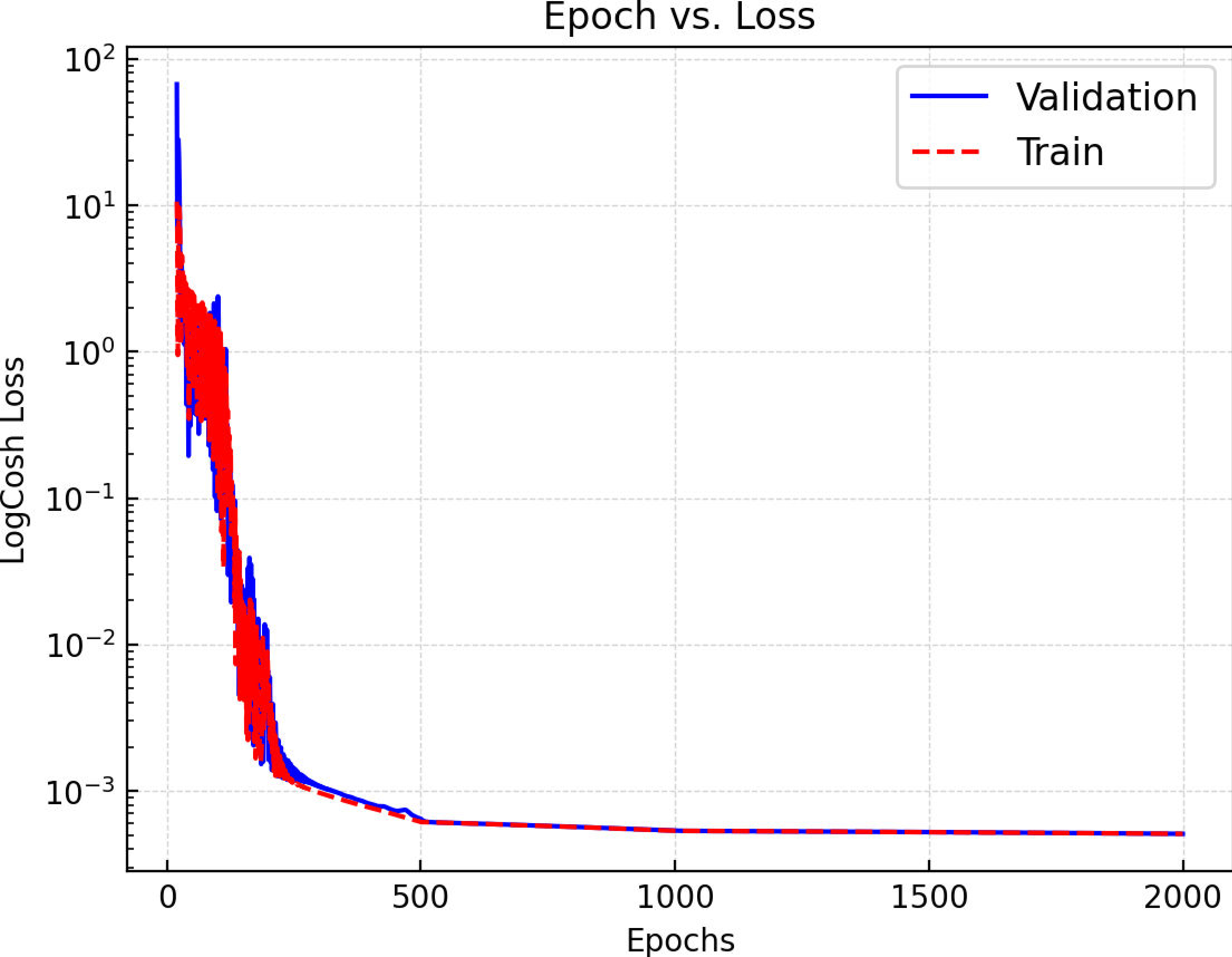}
			\caption{Loss function evolution for each network as a function of the number of epochs. The dark blue solid line indicates the training set loss function evolution and the red dashed line indicates the validation set loss function evolution.}
			\label{fig:loss}
		\end{figure}

		Table~\ref{tab:unet} demonstrates the specific hyperparameters used in the network. The analysis employed the Adam optimizer with standard TensorFlow parameters~\citep{Kingma2014AdamAM,Sashank:2019abs}, and set $\beta_1 = 0.9$ and $\beta_2 = 0.999$. These hyperparameters were meticulously tuned to optimize the network. Ultimately, the optimal values for the initial number of convolutional filters, and batch size were set to $32$ and $2$, respectively. The total number of parameters in this network is $9.04\times10^6$, with a total of $7.4\times10^6$ trainable parameters. Considering the presence of positive and negative values induced by temperature fluctuations in HI data, we selected the ReLU and LeakyReLU activation functions with an alpha value of 1.0 to address this relationship. Figure~\ref{fig:loss} illustrates the evolution of the loss function throughout the training process.

		We employed four optimizers to implement the weight decay strategy, as illustrated in Table~\ref{tab:unet}. The selected decay rate was determined as $10^{-5}$. Furthermore, considering the large dimensions of the images, we had to reduce the batch size to $2$ and $4$. In addition, we compared two optimizers and ultimately selected Adam as the preferred option. We trained for 20000 epochs using a piecewise constant decay learning rate. The specific learning rate decay is set as follows: $\mathbf{boundaries} = [1000, 2000, 4000, 8000, 14000]$; $\mathbf{values} = [0.1, 0.01, 0.001, 0.0005, 0.0001, 0.00005]$. This means that the learning rate is $0.1$ for epochs $0\sim 1000$, $0.01$ for epochs $1000\sim 2000$, and so on. We conducted all the experiments using the {\tt{TensorFlow2}} on NVIDIA A40.

		\begin{figure*}
			\centering
			\includegraphics[width=0.6\textwidth]{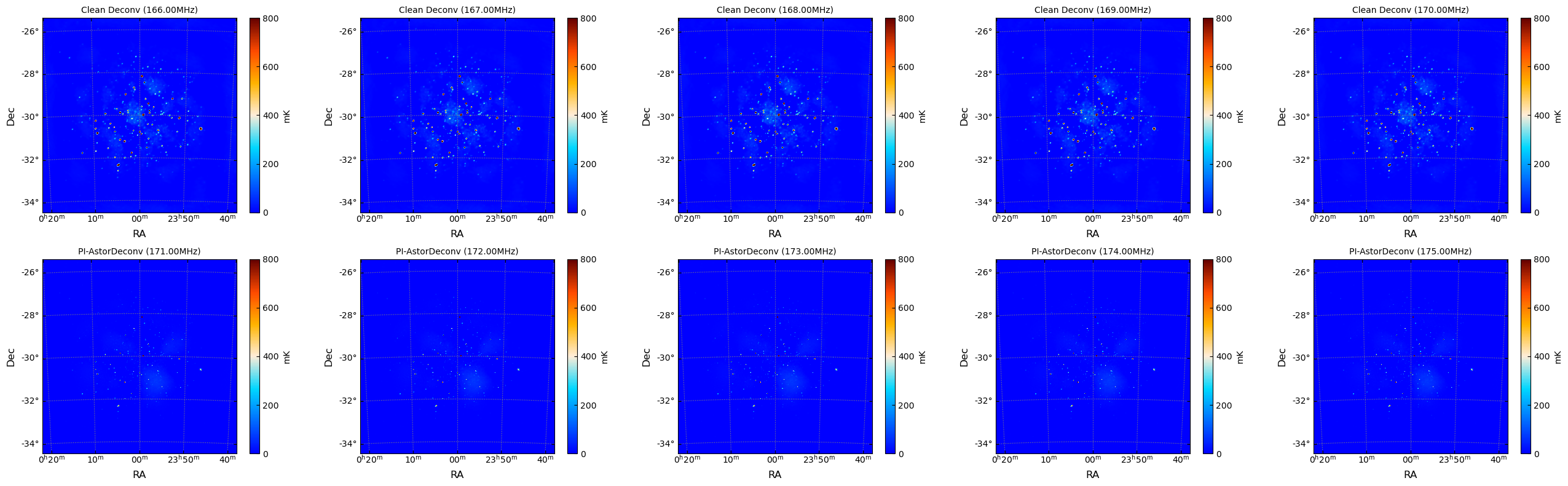}
			\caption{Comparison of the results of the {\tt{CLEAN}} algorithm and the {\tt{PI-AstorDeconv}} algorithm. The frequency increases progressively from $166~\mathrm{MHz}$ to $170~\mathrm{MHz}$. The left column shows five images processed using the {\tt{CLEAN}} algorithm, each of which has undergone $10^6$ deconvolution iterations. The five images in the right column show the processing results of the {\tt{PI-AstroDeconv}} architecture. To facilitate detailed comparison, both columns of images include enlarged illustrations of the brightest sources in each image.}
			\label{fig:result_deconv}
		\end{figure*}

		\section{Results and Discussion}\label{sec:results}
		
		In the previous chapter, we explored in-depth the techniques of interferometric imaging, traditional deconvolution algorithms, and deep learning-based deconvolution methods. In this section, we will conduct a comprehensive and detailed analysis of 21-cm power spectra of the obtained results.

        \subsection{Image Analysis}
        To objectively evaluate the differences in deconvolution effects between  {\tt{CLEAN}}  and  {\tt{PI-AstroDeconv}} , we present five sets of images processed by both algorithms in Figure~\ref{fig:result_deconv}. Each set of images includes results from both algorithms, facilitating a clear and direct comparison of their distinctions and respective advantages. Additionally, to more precisely illustrate the differences in the results, we have magnified the brightest source in each image, allowing for a detailed and nuanced comparison of the differences between these two algorithms.
        
		As shown in Figure~\ref{fig:result_deconv}, the results processed by the {\tt{CLEAN}}  algorithm exhibit noticeably larger point sources compared to those obtained using the {\tt{PI-AstroDeconv}} architecture, with more pronounced nebulous structures surrounding them. To further evaluate the performance of these two algorithms, we will employ Principal Component Analysis ({\tt{PCA}}) for foreground subtraction, followed by a comparative analysis based on this operation, and subsequently conduct a comparative analysis of their 21-D power spectrum based on this operation.

		\subsection{{\tt{PCA}} Foreground Subtraction}

		{\tt{PCA}} is a widely used statistical technique for dimensionality reduction and feature extraction. Its core mechanism involves identifying and retaining the principal directions of variance within the data, known as principal components. By performing an orthogonal transformation, {\tt{PCA}} maps the original data to a new coordinate system where the variance is maximized. When dealing with HI data, foreground signals often exhibit smooth variations across multiple frequency bands, whereas cosmic background signals primarily manifest as variations on larger spatial scales. By calculating the covariance matrix and performing eigenvalue decomposition, {\tt{PCA}} can effectively identify dominant patterns associated with foreground signals and subsequently remove these patterns from the observational data~\citep{deOliveira-Costa:2008cxd,Alonso:2014dhk,Makinen:2020gvh}. Consequently, {\tt{PCA}} is frequently employed as a key tool in the foreground removal analysis for HI, enabling the extraction of the cosmic HI signal by stripping away foreground interference from complex astronomical images.

		The advantage of this method is that it does not require prior knowledge of the specific form of the foreground, making it an effective blind noise reduction technique. However, since the cosmic signal may also exhibit smooth structures on large scales, {\tt{PCA}} might remove cosmological clustering information valuable for research. Therefore, when applying {\tt{PCA}} to remove foregrounds, it is necessary to carefully select the number of principal components to remove, in order to balance the trade-off between noise reduction and the preservation of useful signals.
		
		In another study, we learned that the beam effect can significantly influence the performance of the {\tt{PCA}} algorithm in removing the HI foreground. Therefore, in this research, we first subjected the data to visualization processes and obtained results through imaging. Subsequently, we selected two deconvolution methods for comparison: the traditional algorithm and the {\tt{PI-AstroDeconv}} architecture. The adoption of the advanced deconvolution method stems from the fact that the beam effect of the interferometer array can impact the efficacy of {\tt{PCA}}. Our ultimate goal is to compare the performance of the traditional algorithm and the {\tt{PI-AstroDeconv}} architecture in mitigating the beam effect.

		\subsection{21-cm Power Spectrum Analysis}

		To ensure that the correction of the beam effect during the data analysis process did not compromise the authenticity of the cosmic matter distribution, we conducted a 21-cm power spectrum analysis on the processed data. By thoroughly analyzing the 21-cm power spectrum, we can assess the effectiveness of the beam effect correction and determine whether crucial physical information has been preserved.
		
		The 21-cm power spectrum is a statistical tool used to measure the spatial correlation of HI in two different directions. Specifically, it focuses on the distribution of HI parallel to the line of sight ($k_\parallel$) and perpendicular to the line of sight ($k_\perp$). By analyzing the 21-cm power spectrum, we can investigate the early state of the universe, which is one of the objectives of the SDC3a project, namely, to probe the EoR of the universe through observations of the HI. Additionally, the 21-cm power spectrum can also be used to study the distribution patterns of galaxies and galaxy clusters in the universe, revealing the characteristics of HI gas distribution within galactic disks. And the distribution of HI is influenced by the gravitational potential of dark matter, the analysis of the power spectrum allows for the indirect detection and study of the properties of dark matter.

         We have performed {\tt{PCA}} on the observational data pr-eprocessed by both {\tt{WSClean}} and {\tt{PI-AstroDeconv}}, setting the number of principal components to 12 during the {\tt{PCA}} process. The residual signals obtained after {\tt{PCA}} processing were considered as the foreground-cleaned signals. Finally, we conducted a 21-cm power spectrum analysis on the data cubes using the $\tt{tools21cm}$~\footnote{https://tools21cm.readthedocs.io/} package.

\begin{figure*}
			\centering
			\includegraphics[width=1\textwidth]{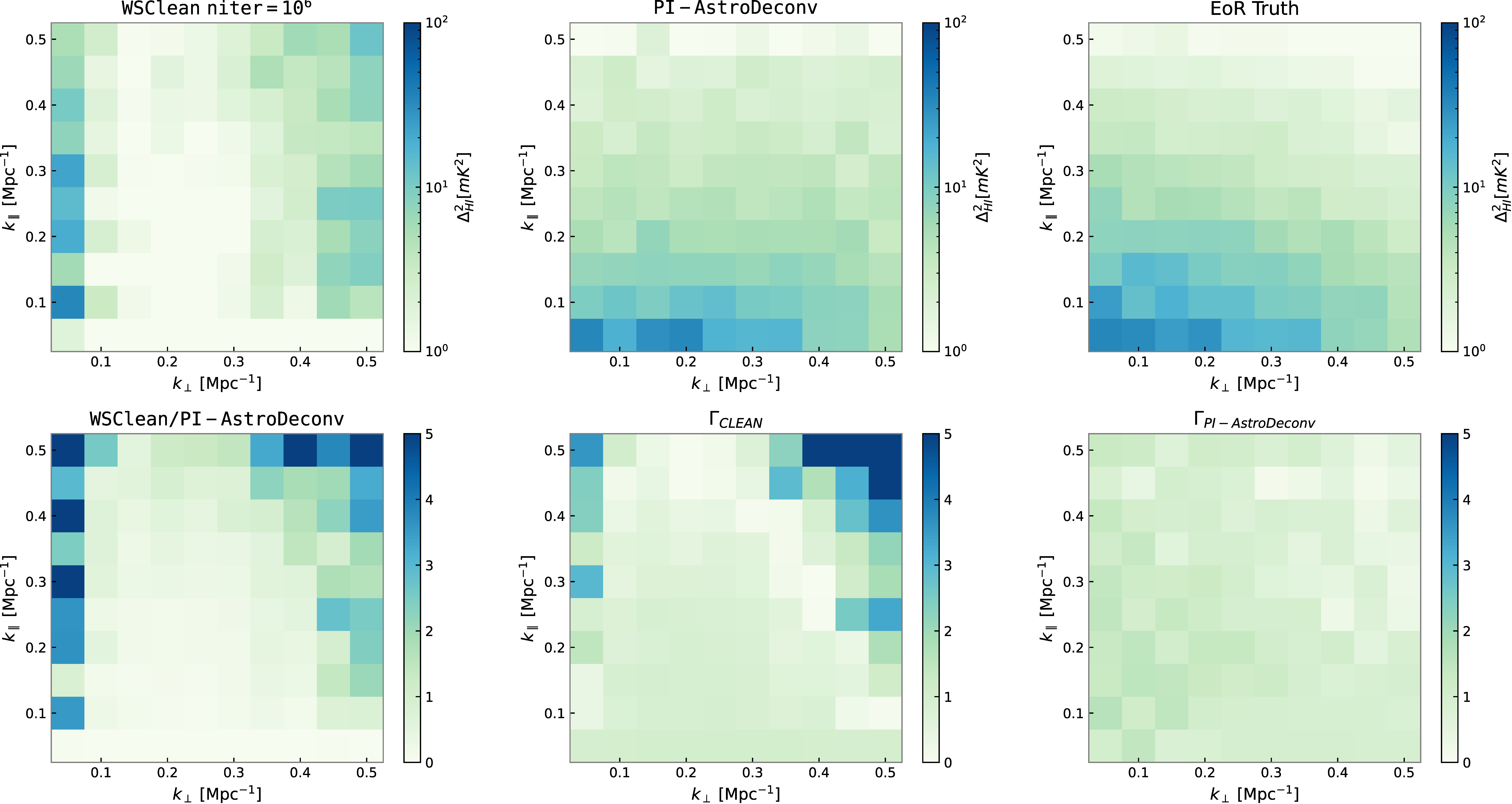}
			\caption{The 21-cm power spectrum of the {\tt{PCA}} cleaned map, and the ratios between them. Top-left panel: The residual power spectrum after correcting for the beam effect using the {\tt{CLEAN}} algorithm~(niter=$10^6$) and removing foreground contamination with the {\tt{PCA}} algorithm. Top-middle panel: The residual power spectrum after correcting for the beam effect using the {\tt{PI-AstroDeconv}} architecture and removing foreground contamination with the {\tt{PCA}} algorithm. Top-right panel: The true EoR  power spectrum without any foreground or instrumental effects. Bottom-left panel: The ratio between the power spectra shown in the top-left and top-right panels. Bottom-middle panel: The $\Gamma_{\tt{CLEAN}}$ value calculated based on Equation~(\ref{eq:2dpsratio}). Bottom-right panel: The $\Gamma_{\tt{PI-AstroDeconv}}$ value calculated based on Equation~(\ref{eq:2dpsratio}).}
			\label{fig:power_spectrum_two}
		\end{figure*}

         In the upper part of Figure~\ref{fig:power_spectrum_two}, we present the 21-cm power spectra processed using the {\tt{CLEAN}} algorithm and the {\tt{PI-AstroDeconv}} architecture, as well as the ideal EoR signal power spectrum. Specifically, the top-left panel shows the residual power spectrum after correcting for the beam effect and removing foreground contamination using the {\tt{CLEAN}} algorithm; the top-middle panel displays the results processed with the {\tt{PI-AstroDeconv}} architecture following the same procedure; and the top-right panel provides the ideal EoR signal power spectrum, free from any foreground or instrumental effects.

         The analysis of the discrepancy plot indicates that, compared to other positions, there is a significant difference between the results processed with {\tt{WSClean}} and the true signals, to the extent that it almost fails to show the trend of HI distribution. However, the results obtained through {\tt{PI-AstroDeconv}}  processing are in good agreement with the true signals on large scales (i.e., in regions of small $k$ values). On small scales (i.e., at large $k$ values), the values predicted by {\tt{PI-AstroDeconv}}  are relatively high, which may be due to the {\tt{PCA}} algorithm overfitting the large-scale structures, leading to partial removal or residual foreground contamination of small-scale signals. Although there are some deviations between the residuals after {\tt{PCA}} processing and the true signals, our focus is on evaluating the improvement of {\tt{PI-AstroDeconv}}  over the {\tt{CLEAN}} algorithm in terms of beam effect correction capabilities. This phenomenon does not affect our conclusion that {\tt{PI-AstroDeconv}}  can effectively enhance the ability of the {\tt{CLEAN}} algorithm to remove convolution effects.
		
		To facilitate a direct comparison between the two, we have defined a formula to highlight these differences:
		\begin{equation}\label{eq:2dpsratio}
			R(k_\parallel, k_\perp) = |\frac{P_{\rm pred}(k_\parallel, k_\perp)}{P_{\rm truth}(k_\parallel, k_\perp)} - 1|,
		\end{equation}
		where $P_{\rm pred}(k_\parallel, k_\perp)$ is the predicted power spectrum after processing with {\tt{PCA}}, and $P_{\rm truth}(k_\parallel, k_\perp)$ is the true power spectrum. 

        Similarly, in the lower half of Figure~\ref{fig:power_spectrum_two}, we present the ratios between the 21-cm power spectra and their corresponding $\Gamma$ values. The bottom-left panel shows the ratio of the power spectrum after correcting for the beam effect using both {\tt{WSClean}} and {\tt{PI-AstroDeconv}}, combined with {\tt{PCA}} for foreground removal. The bottom-middle panel displays the result calculated based on Equation~(\ref{eq:2dpsratio}) using {\tt{WSClean}}, while the bottom-right panel presents the result obtained using the {\tt{PI-AstroDeconv}} architecture with the same equation.

		We also have simplified the quantification process, as detailed in the following formula:
		\begin{equation}\label{equ:acc}
			\Gamma= {\rm mean}\left\{R(k_\parallel, k_\perp)_{\rm X}\right\},
		\end{equation}
		where the notation ${\rm mean}\{\}$ is used to denote the operation of calculating the average of the parameter $R(k_\parallel, k_\perp)$ and $\rm X$ represents {\tt{WSClean}} and {\tt{PI-AstroDeconv}}, respectively. Through our calculations, we obtained a result of $\Gamma_{\tt{WSClean}} = 3.472$ and $\Gamma_{\tt{PI-AstroDecovn}} = 0.756$, which demonstrates that the method we utilized has effectively achieved beam mitigation to a notable degree.

		\section{Conclusion}\label{sec:conclusion}
		
		The beam effect is an image distortion in astronomical observations induced by factors such as the optical characteristics of telescopes. Traditional deconvolution methods, like the {\tt{CLEAN}} algorithm, rely on complex models and prior knowledge, thus having limited adaptability. This paper presents a physics-informed unsupervised deep learning architecture for beam deconvolution, termed {\tt{PI-AstroDeconv}}, which can learn beam characteristics from raw data without requiring a physical model. Additionally, the powerful nonlinear modeling capability of the deep learning model effectively restores image details.

        We conducted a quantitative comparison of the results. By introducing a parameter $\Gamma$ that measures the discrepancy between the estimated and the true power spectra, 
        we found that $\Gamma_{\tt{PI-AstroDeconv}}=0.756$ significantly outperforms $\Gamma_{\tt{CLEAN}}=3.472$, which corresponds to the {\tt{CLEAN}} algorithm. This indicates that our proposed method demonstrates a clear advantage in terms of reconstruction accuracy.

        Furthermore, the adoption of unsupervised architecture effectively reduces the reliance on expert knowledge, making astronomical data processing more efficient and automated. This characteristic is particularly important for the rapid processing and real-time analysis of observational data, especially in large-scale survey scenarios involving massive datasets. Although the proposed architecture requires separate training at different frequencies, the method can be adapted to other telescope systems by simply modifying the shape of the telescope beam (as shown in Figure~\ref{fig:deconv_unet})~\citep{ni2024pi}. Therefore, the method exhibits high adaptability and can be readily transferred to different observational conditions and instrument configurations, optimizing resource utilization in data processing.
		
		It is worth mentioning that {\tt{PI-AstroDeconv}} has good generalization ability and can be applied to other telescope systems, such as Hubble and Webb Telescope, to provide astronomers with clearer and more detailed images of the universe~\citep{ni2024pi}. This will not only help to more accurately measure astrophysical parameters such as star brightness and galaxy morphology, but also may reveal astronomical phenomena that were difficult to capture by previous observation methods, thus promoting new scientific discoveries.

		\section*{Acknowledgements}
        
		This work was supported by National Key R\&D Program of China (No. 2023YFE0110500), by the Leading Innovation and Entrepreneurship Team of Zhejiang Province of China (No. 2023R01008), by Zhejiang Provincial Natural Science Foundation of China (No. LY24A030001), and by Key R\&D Program of Zhejiang (No. 2024SSYS0012).



\bibliography{ref}{}
\bibliographystyle{aasjournal}



\end{document}